\documentclass[aps,prl,twocolumn,groupedaddress,showpacs,amsmath,amssymb]{revtex4}

\usepackage{bm} \usepackage{amsmath} \usepackage{amssymb} \usepackage{latexsym} \usepackage{amsfonts} \usepackage{graphicx}

\begin{document}

\title{Momentum exchange between light and a single atom: Abraham or Minkowski?}

\author{E. A. Hinds}

\affiliation{Centre for Cold Matter, Imperial College, Prince
Consort Road, London SW7 2AZ, United Kingdom}

\author{Stephen M. Barnett}

\affiliation{SUPA, Department of Physics, University of Strathclyde, Glasgow G4 0NG, United Kingdom}

\date{\today}

\begin{abstract}
We consider the forces exerted by a pulse of plane-wave light on a single atom. The leading edge of the pulse exerts a dispersive force on the atom, and this modifies the atomic momentum while the atom is enveloped in the light. The standard view of the optical dipole force indicates that red-detuned light should attract the atom towards high intensity. This should increase the average momentum per photon to $\textbf{p}_{0} n$, where $\textbf{p}_{0}$ is the photon momentum in free space and $n$ is the average refractive index due to the presence of the atom in the light. We show, however, that this is the wrong conclusion and that the atom is in fact repelled from the light by the dispersive forces, giving the photons a momentum $\textbf{p}_{0} /n$. This leads us to identify Abraham's optical momentum with the kinetic momentum transfer.  The form due to Minkowski is similarly associated with the canonical momentum. We consider the possibility of demonstrating this in the laboratory, and we note an unexpected connection with the Aharonov-Casher effect.
\end{abstract}

\maketitle

The study of light within a medium has identified a surprising number of candidates for the density of optical momentum.
Principal among these are the forms given by Minkowski, $\textbf{S}_\textrm{Min}=\textbf{D}\times\textbf{B}$, and by Abraham, $\textbf{S}_\textrm{Abr}=\textbf{E}\times\textbf{H}/c^2$ \cite{Brevik}.  When integrated over the volume of
the medium these alternatives ascribe different momenta to a photon propagating in the material: $\textbf{p}_\textrm{Min}=\textbf{p}_{0}n$ and $\textbf{p}_\textrm{Abr}=\textbf{p}_{0}/n$ respectively, where
$\textbf{p}_{0}$ is the free-space photon momentum and $n$ is the refractive index.  It is intriguing to note that
the uniform motion of the centre of mass-energy leads, unambiguously, to the Abraham expression \cite{Brevik}.
Consideration of diffraction, however, leads equally convincingly to the Minkowski form \cite{Miles}.  There is,
moreover, a bewildering array of experimental studies and associated theoretical analyses which appear to favor
one or other of these momenta or, indeed, others \cite{Brevik,Lots,Gordon73}.
Global momentum conservation is certainly not in doubt, and it is clear that both the Minkowski and Abraham forms
are true momentum densities, but understanding which provides the natural description of any given phenomenon remains
a challenge. At a fundamental level we can trace the
origin of this problem to the difficulty in separating the electromagnetic field from the matter \cite{RodSteve,HalinaReview07}.  The absence of a unique optical momentum has led some authors
to concentrate first on the Lorentz force and to calculate from this the relevant momentum transfer
\cite{RodSteve,Gordon73,Rodney,Masud,Barnett06}.

In this letter, we consider light interacting with a single atom. This medium is simple enough for us to identify the optical momentum with some clarity.  We find that \emph{both} the Minkowski and Abraham momenta have
readily identifiable roles associated, respectively, with the canonical and kinetic momenta of the atom.
\begin{figure}[t]
\includegraphics{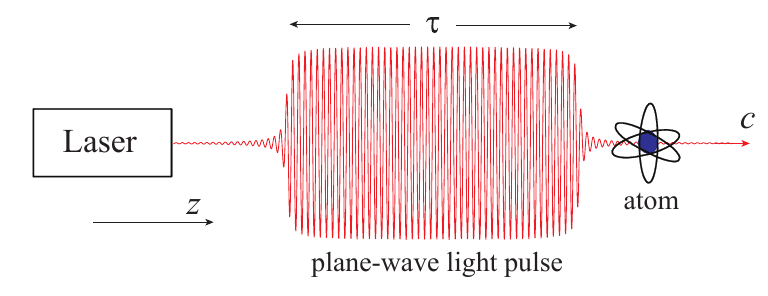}
\caption{\label{fig:expt}Illustration of the system under consideration. }
\end{figure}

 Let us consider a single atom with an electric-dipole moment ${\bf d}$ interacting with a light pulse.
Within the dipole approximation the polarization is localized at the position of the atom,
${\bf r}_{\rm atom}$, and our electric displacement and magnetic flux densities are simply
$\textbf{D}=\epsilon_0\textbf{E}+\textbf{d}\,\delta(\textbf{r}-\textbf{r}_{\textrm{atom}})$ and
${\bf B} = \mu_0{\bf H}$.  It follows that there is a very simple relationship between the Minkowski and
Abraham momenta:
\begin{equation}\label{eq:lightMomenta}
\int\textbf{S}_\textrm{Min}d^{3}r=\int\textbf{S}_\textrm{Abr}d^{3}r+\textbf{d}\times\textbf{B}
({\bf r}_{\rm atom}).
\end{equation}
In order to understand the nature of this difference let us calculate the force exerted on our atom by a plane-wave
pulse of light passing the atom, as depicted in Fig.~\ref{fig:expt}.  The $i^{th}$ component of the Lorentz force
exerted on our slowly moving atom ($v \ll c$) is simply \cite{Gordon73,Barnett06}
\begin{equation}
\label{eq:Barnettforce}
F_i = \left({\bf d}\cdot\nabla\right)E_i + \left(\dot{\bf d}\times{\bf B}\right)_i .
\end{equation}
Alternatively, the Maxwell equation $\dot{\bf B} = -\nabla\times{\bf E}$ allows us to rewrite this in the form
\cite{Gordon73,Barnett06}
\begin{equation}
\label{eq:Gordonforce}
F_i=\mathcal{F}_{1i}+\mathcal{F}_{2i}=\textbf{d}\cdot\frac{\partial}{\partial x_{i}}\textbf{E}
+\frac{\partial}{\partial t}(\textbf{d}\times\textbf{B})_{i} .
\end{equation}
In the literature of laser cooling and trapping, the force is normally given as just $\boldmath{\mathcal{F}}_{1}$, without the term $\frac{\partial}{\partial t}(\textbf{d}\times\textbf{B})$ (Refs.~\cite{G&A80,CohensError} are two examples of many). For many practical situations the missing term is of little consequence.  For us, however, it is not only significant but it accounts fully for the difference between the Abraham and Minkowski momenta.

The plane-wave laser field traveling along the $z$-axis can be written as
\begin{equation}\label{eq:Efield}
\textbf{E}=\mathcal{E}(\omega t-k z)\cos\left(\omega t-k z\right)\hat{\textbf{e}}\,,
\end{equation}
where $\hat{\textbf{e}}$ is the constant, transverse polarisation and $k$ is the wavevector. This gives the expectation value of $\mathcal{F}_{1z}$ as
\begin{equation}\label{eq:GordonF1}
\langle \mathcal{F}_{1z}\rangle=\langle\textbf{d}\cdot\hat{\textbf{e}}\rangle\left(\frac{\partial\mathcal{E}}{\partial z}\cos\left(\omega t-k z\right)+k\mathcal{E}\sin\left(\omega t-k z\right)\right)\,.
\end{equation}
The factor $\langle\textbf{d}\cdot\hat{\textbf{e}}\rangle$ is the expectation value of the component of electric dipole moment that couples to the field. The rest of the right hand side is the field gradient. This has been taken outside the integral over the atomic wavefunction under the assumption that the centre of mass wavepacket is small compared with the distance over which the field gradient varies. This is a good approximation in nearly all cases because the size of the wavepacket is smaller than the optical wavelength for any temperature above the recoil limit.

The dipole $\langle\textbf{d}\cdot\hat{\textbf{e}}\rangle$ can be obtained from the optical Bloch equations that describe the evolution of the atomic density matrix in the presence of the light field. For a 2-level atom within the rotating-wave
approximation, this has a steady-state solution
\begin{equation}\label{eq:BlochDipole}
\langle\textbf{d}\cdot\hat{\textbf{e}}\rangle=2 D\left(u \cos\left(\omega t-k z\right)-v\sin\left(\omega t-k z\right)\right)\,,
\end{equation}
where $D$ is the off-diagonal matrix element of $\textbf{d}\cdot\hat{\textbf{e}}$ and
\begin{equation}\label{Eq:uv}
\left(\begin{array}{l}
u\\
v\\
\end{array}\right) =
\left(\begin{array}{l}
\delta\\
\gamma\\
\end{array}\right)\frac{\frac{1}{2}\Omega}{\delta^2+\gamma^2+\frac{1}{2}\Omega^2}\,
\end{equation}
describe the components of the driven dipole in phase and in quadrature with the driving field \cite{Cohens Book,Allen}. Here $\delta=\omega-\omega_{\textrm{at}}$ is the detuning of the light frequency from the atomic transition frequency, $\gamma$ is half the spontaneous decay rate of the population in the upper state and $\Omega$ is the Rabi frequency, defined by $\hbar\Omega=-D\mathcal{E}$.

On substituting Eqs.~(\ref{eq:BlochDipole}) and (\ref{Eq:uv}) into Eq.~(\ref{eq:GordonF1}) and taking the time average over an optical cycle, we obtain the average force
\begin{equation}\label{eq:AvgGordonF1}
\overline{\langle \mathcal{F}_{1z}\rangle}=D\left(u\,\nabla\mathcal{E}-v\,k\mathcal{E}\right)\,.
\end{equation}
The first term, known as the optical dipole force or gradient force, has a dispersive frequency dependence and is associated with the in-phase part of the driven dipole. The second term, the scattering force, has an absorptive frequency dependence and is due to the scattering of momentum out of the light beam by spontaneous emission.

We shall return later to the scattering force, but let us consider first the gradient force in Eq.~(\ref{eq:AvgGordonF1}) and assume that the detuning is red, so that $\delta$ is negative. Then the leading edge of the pulse exerts a gradient force that attracts the atom towards higher light intensity. Once the atom is fully enveloped by the pulse, the net momentum imparted to it by this gradient force is
\begin{eqnarray}\label{Eq:ODimpulse}
P_{\nabla \mathcal{E}}&=&\int D u \nabla \mathcal{E} dt\nonumber\\
&=&\frac{\hbar \delta}{2c}\ln\left(1+\frac{\frac{1}{2}\Omega_{0}^2}{\delta^2+\gamma^2}\right)\simeq-\frac{1}{2}D u \mathcal{E}_{0}/c\,.
\end{eqnarray}
Here $\Omega_{0}$ is the Rabi frequency due to the electric field inside the light pulse. This result follows from the replacement of $dt$ by $dz/c$, which is satisfactory because the velocity of the atom is negligible in comparison with $c$. We also neglect in this formula the very small difference between $c$ and the group velocity of the light in the presence of the atom. The final expression on the right of Eq.~(\ref{Eq:ODimpulse}), applies in the linear response approximation, where $\Omega^2\ll\delta^2+\gamma^2$.  This is reasonable, of course, because the Minkowski and Abraham momenta differ in their
dependences on the (linear) refractive index.

We turn now to $\mathcal{F}_{2}$, the second term on the right hand side of Eq.~(\ref{eq:Gordonforce}), which would normally be neglected. This also acts throughout the leading edge of the pulse, giving the atom an additional momentum along the $z$ direction of
\begin{equation}\label{Eq:ODimpulseA}
P_{d\times B}=\langle\textbf{d}\cdot\hat{\textbf{e}}\rangle \mathcal{E}_0/c \simeq D u \mathcal{E}_{0}/c\,.
\end{equation}
Here we have used Eq.~(\ref{eq:BlochDipole}) for the expectation value of the dipole and in the last step we have again averaged over an optical cycle. We see that the impulse $\textbf{d}\times\textbf{B}$ is \emph{twice as large} as that generated by the gradient force and is in the opposite direction. Consequently, the total momentum imparted to the atom by the dispersive force acting on the leading edge of the pulse is equal to that of the gradient force, but in the opposite direction: with red detuning the atom is repelled from the light, not attracted to it.

Why do we normally neglect the impulse $\textbf{d}\times\textbf{B}$ when it is so significant here? The reason is that this impulse depends on the change in field strength but not on the time taken for the atom to enter the field. By contrast, the gradient force produces an impulse proportional to the time for which the force acts. In many applications, the intensity distribution is static and then this time is set by the velocity of the atom, not by the speed of light. In those cases, the impulse generated by the gradient force is much larger than $\textbf{d}\times\textbf{B}$, which can then be safely neglected.

In the discussion above, we use Eq.~(\ref{eq:Gordonforce}) to describe the force because this is closer to the expression ($\mathcal{F}_{1}$) that is normally used in the laser cooling and trapping literature. If instead we use Eq.~(\ref{eq:Barnettforce}), we obtain the same result, but with a physically more intuitive description. In this formulation, the first term derives from the Coulomb force, and because the electric field has no component in the $z$ direction this term makes no contribution to the force in the system under consideration. Instead, the force comes entirely from the second term in Eq.~(\ref{eq:Barnettforce}), which can be immediately understood as a manifestation of the magnetic Lorentz force $q\textbf{v}\times\textbf{B}$. On substituting Eqs.~(\ref{eq:BlochDipole}) and (\ref{Eq:uv}) into Eq.~(\ref{eq:Barnettforce}) and taking the time average over an optical cycle, we obtain the average force
\begin{equation}\label{eq:AvgBarnettF}
\overline{\langle F_z \rangle}=D \left(\frac{\partial u}{\partial t}\frac{\mathcal{E}}{c}- v k\mathcal{E} \right)\,.
\end{equation}
The first term is the dispersive part of this force and integrating it over the leading edge of the pulse, we obtain once again an atomic momentum of $+D\mathcal{E}_{0}u/(2c)$, indicating repulsion from the red-detuned light. The second term is again the scattering force that we have already seen in Eq.~(\ref{eq:AvgGordonF1}).

Knowing that the dispersive force gives the atom additional momentum in the $z$ direction, it follows from the conservation of momentum that this same interaction must \emph{reduce} the momentum of the light by a corresponding amount.  As the number of photons in a pulse of volume $V$ is $\tfrac{1}{2}\epsilon _{0}\mathcal{E}_{0}^2 V/(\hbar \omega)$, the momentum of each photon becomes
\begin{equation}\label{eq:pgamma}
{\bf p}={\bf p}_0\left(1- \frac{D u}{\epsilon_{0} \mathcal{E}_{0} V} \right)\simeq\frac{{\bf p}_0}{1+ D u/(\epsilon_{0} \mathcal{E}_{0} V) }\,,
\end{equation}
the last step being valid because the correction is small. From Eq.~(\ref{eq:BlochDipole}) we see that the real (i.e. in-phase) part of the electric susceptibility is $\chi\prime=2 D u/(\epsilon_{0} \mathcal{E}_{0} V) $.  This is small, of course, so the real part of the refractive index is well approximated by $1+\tfrac{1}{2}\chi\prime$ and therefore
\begin{equation}\label{eq:pgamma-n}
{\bf p}=\frac{{\bf p}_0}{n}\,.
\end{equation}
This result has again assumed that $\Omega^2\ll\delta^2+\gamma^2$, so that the linear response approximation is valid.
We see that the momentum transfer to the atom requires us to identify the Abraham momentum with each photon in our pulse.

We note that if the term $\mathcal{F}_2$ in Eq.~(\ref{eq:Gordonforce}) is neglected, then the change of photon momentum due to the atom is exactly reversed and we obtain instead the result ${\bf p}={\bf p}_0 n$, which is the Minkowski result. This can be understood directly from Eq,~(\ref{eq:lightMomenta}) because dropping $\mathcal{F}_2$ from the force on the atom, adds $\textbf{d}\times\textbf{B}$ to the momentum of the light, thereby converting the Abraham momentum into the Minkowski momentum.

If the momentum transfered by the Lorentz force has the Abraham value, then what is the physical significance of the
Minkowski momentum?  To understand this we note that the total momentum is conserved and that this can be written in
two different ways:
\begin{equation}
\label{TotMom}
M\dot{\bf r}_{\rm atom} + \int\textbf{S}_\textrm{Abr}d^{3}r = {\bf p}_{\rm atom} +
\int\textbf{S}_\textrm{Min}d^{3}r ,
\end{equation}
where $M$ is the mass of the atom and ${\bf p}_{\rm atom} = M\dot{\bf r}_{\rm atom}
- \textbf{d}\times\textbf{B}({\bf r}_{\rm atom})$ is the \emph{canonical} momentum of the atom \cite{Lembessis}. If,
as our analysis has suggested, the Abraham momentum is associated with the \emph{kinetic} momentum exchanged between
the atom and the field, then it follows that the Minkowski momentum must similarly be associated with the
\emph{canonical} momentum.  In quantum theory it is the canonical momentum that is associated with the wavlength and
it may be for this reason that it is the Minkowski momentum that gives the simplest description of wave-like
phenomena, such as diffraction \cite{Miles}, while the Abraham momentum is naturally associated with particle-like
phenomena associated with forces and the kinetic momentum.

In an experiment to test the kinetic momentum transfer, the atom would also experience the scattering force $F_\textrm{scatt}$, which appears as $-D v k \mathcal{E}$ in either formulation of the electric dipole force, as seen in  Eqs.~(\ref{eq:AvgGordonF1}) and (\ref{eq:AvgBarnettF}). The proportionality to $v$ alerts us to the fact that this force comes from the dissipative interaction between the field and the component of the dipole in quadrature with it. Using Eq.~(\ref{eq:BlochDipole}), we can re-write this force as
\begin{equation}\label{eq:scatterForce}
{\bf F}_\textrm{scatt}=
{\bf p}_0 \left(\frac{ \tfrac{1}{2}\Omega^{2}}{\delta^2+\gamma^2+\tfrac{1}{2}\Omega^{2}}\gamma \right)\,.
\end{equation}
From the upper level population in the steady-state of the optical Bloch equations, one identifies the factor in parentheses as the rate of spontaneous scattering by the atom, showing this force to be the momentum removed per second from the incident beam by scattering light into random directions. In other words, $F_\textrm{scatt}$ is the force due to absorption. In the presence of this force it is practical to ask whether an experiment could discern the effect of the dispersive force in order to discriminate between the Abraham and Minkowski results and test the theory presented here. When the pulse leaves the atom behind, the dispersive force on the trailing edge removes the momentum imparted by the leading edge. The final momentum of the atom therefore measures only the scattering force. There is however a displacement, which has contributions from both types of force.

Immediately after the trailing edge of a pulse of duration $\tau$, the two displacements are in the ratio
\begin{equation}\label{eq:displacement ratio}
\frac{\Delta x_\textrm{dispersion}}{\Delta x_\textrm{absorption}}=\frac{D u \mathcal{E}_0/c}{D v \mathcal{E}_0 k \tau}=\frac{\delta}{\omega}\frac{1}{\gamma \tau}\,.
\end{equation}
The pulse duration must be long in comparison with $\gamma$ for our treatment to be valid because we neglect the transient response of the atom to the changes in the field. Shorter pulses can certainly be produced in the laboratory, but the concept of refractive index as used in the Abraham/Minkowski discussion assumes that the response is proportional to the field and does not depend on its history.  We therefore take it that $\gamma \tau\gg1$. We can also assume that $|\delta|\lesssim\omega$. This is rigorously true for red detuning, but it is also true in practice for blue detuning because light detuned from resonance by more than the transition frequency would have to be unfeasibly intense to generate a measurable effect and would also be likely to couple more strongly to higher dipole excitations of the atom. It would seem, therefore, that both the momentum and the displacement of the atom are substantially dominated by the effects of radiation pressure, making this an unsuitable way to determine the sign of the dispersive force on the atom. However, it should be possible using the method of slow light \cite{Hau99} to reduce the group velocity of the light to give the gradient force a much longer time to operate, thereby enhancing the effect of interest \cite{Patrik}. At the same time this method can reduce the undesirable spontaneous scattering rate.

We conclude by noting that the Minkowski and Abraham momenta differ also in their dependence on the magnetic field.  If
our atom has a magnetic-dipole moment $\bf m$, then we have ${\bf B} = \mu_0[{\bf H} + {\bf m}\delta({\bf r} -
{\bf r}_{\rm atom})]$ and we can follow the above analysis to find that the Minkowski and Abraham momenta differ by
${\bf m}\times{\bf E}/c^2$.  As with an electric dipole, it is the Abraham momentum that is associated with the kinetic
momentum transfer.  The difference between this and the canonical momentum, associated with the Minkowski momentum, is
precisely the interaction responsible for the Aharanov-Casher effect \cite{Aharonov}.

In summary, we have discussed the momentum transferred by a pulse of light to a single 2-level atom. When the light is tuned to the red side of resonance, the dispersive part of the force repels the atom from the light in contradiction to the usual view that red-detuned light will be attractive. We have shown that the difference arises from a term $\textbf{d}\times\textbf{B}$ in the momentum, which is normally neglected in the literature of laser cooling and trapping. This led us to consider the momentum transferred to the light. We have found that Abraham's optical momentum corresponds to the kinetic momentum transfer, whereas the form due to Minkowski is associated with canonical momentum, the difference between the two being precisely the $\textbf{d}\times\textbf{B}$ term. We then considered whether this momentum transfer might be measured in the laboratory. We found that the effect of the scattering forces always masks that of the dispersive force on a single 2-level atom, but not necessarily in a vapor of many multi-level atoms.  Finally, we noticed that the interaction giving rise to the Aharonov-Casher effect is exactly the magnetic analog of the electric-dipole coupling that causes the Abraham/Minkowski difference discussed here.

\begin{acknowledgments}
This work was supported by the Royal Society, the Wolfson Foundation, the UK EPSRC and the Rank Prize Foundation.
\end{acknowledgments}

\end{document}